\documentclass[usenatbib]{basi}
\usepackage[british]{babel}
\usepackage[varg]{txfonts}
\usepackage{rotating}
\usepackage{dcolumn}
\begin{document}
\title[The enigma of jets and outflows from young stars]{The enigma of jets and outflows from young stars}
\author[D.~Coffey 2011]%
       {D.~Coffey$^{1,2}$\thanks{email: \texttt{dac@cp.dias.ie}}\\
       $^1$I.N.A.F. - Osservatorio di Arcetri, Florence, Italy\\
       $^2$The Dublin Institute for Advanced Studies, Ireland}

\pubyear{2011}
\volume{00}
\pagerange{\pageref{firstpage}--\pageref{lastpage}}

\date{Received \today}

\maketitle
%------------------------------------------------------------------------------%
% abstract and keywords                                                        %
%------------------------------------------------------------------------------%
\label{firstpage}

\begin{abstract}

Research in recent decades has seen many important advances in
understanding the role of jets and outflows in the star formation process.
Although, many open issues still remain, multi-wavelength high resolution
observations have provided unprecedented insights into these bizarre
phenomena. An overview of some of the current research is given, 
in which great strides have been made in addressing 
fundamental questions such as: how are jets generated? what is the jet
acceleration mechanism? how are jets collimated? what is the relationship
between accretion and ejection? how does mass accretion proceed? 
do jets somehow extract angular momentum? 
and finally, is there a universal mechanism for jet generation on all scales from brown
dwarfs to AGNs?

\end{abstract}


\section{Introduction}\label{s:intro}

The striking phenomena of jets and outflows from young stars (Bally et al. 2007; Ray et al. 2007) was wholly unanticipated by theorists, who are still struggling to understand the basic mechanisms involved. Since jets are observed to 
transport signiÞcant amounts of energy and momentum away from the central source, it is 
likely that they play an important role in the evolution of the parent star. 
However, since the jet production mechanism works on very small scales, 
and the central jet engine is often heavily embedded, observations struggle 
to distinguish between currently proposed theoretical models. 
As a result, many open issues remain, including an understanding of how mass accretion proceeds; how excess angular momentum is extracted; the nature of the accretion/ejection relationship; the jet generation mechanism; and whether the process is similar on all masses and length scales. Notwithstanding, recent observational advances have seen exciting progress in addressing these fundamental issues. 

\subsection{Models of jet generation}\label{s:models}

%It is generally acknowledged that magneto-centrifugal forces are responsible for jet launching. 
It is now widely accepted that to make a fast collimated jet requires an accretion disk threaded by a large scale magnetic field. This is because neither gas pressure nor radiation pressure appear up to the task of collimating large momentum flux. These so called magneto-centrifugal models propose that the magnetic fields in the collapsing, rotating cloud core are advected with the accretion flow and so form an hour glass shaped magnetic field as a result of this inward motion of the disk material. 
Matter is picked up from the disk and forced along magnetically dominated 
accretion columns to high stellar latitudes where it is then accreted onto the star.  
Most have adopted this magnetic geometry, although no direct observations have yet determined a typical magnetic field strength/configuration for accretion disks around low mass stars. 

It is proposed that, during this accretion process, magneto-centrifugal forces are responsible for the launch, 
acceleration and collimation of high velocity protostellar jets. 
%Particles are lifted from the star-disk system and accelerated vertically away from the disk-plane like a bead-on-a-wire along the magnetic field lines. The angular velocity of the particles remains unchanged, so long as the ions are forced to move with the field (as a result of the so called Ôfrozen-inÕ field condition assumed in many astrophysical situations). This means that, as the gas increases its distance from the disk-plane, the angular momentum of the gas increases per unit mass. Conservation of angular momentum then implies that the disk must decrease its angular momentum correspondingly. Beyond the Alfv\'{e}n surface (i.e. where the kinetic energy density matches the magnetic energy density), the gas now has enough energy to dominate over the field strength, and so the Ôfrozen-inÕ field is forced to curve in the direction of the gas rotation. The outflow now takes on a rotational velocity, resulting in the introduction of a toroidal field. The resulting Lorentz force is directed towards the rotation axis (i.e. Ôz-pinchÕ or Ôhoop stressÕ), thus collimating the outflow. This fascinating picture appears to provide a natural explanation for the long standing problem of angular momentum dissipation in protostars. 
Currently there are three main steady magnetohydrodynamic (MHD) model contenders, which differ mainly 
in their origin of magnetic forces which drive the jet. 
The first is the {\em Stellar wind} in which the jet launching point is the stellar surface
(e.g. Matt \& Pudritz 2005). However, difficulties remain in achieving sufficient angular momentum extraction to slow the stellar rotation to the observed rate via a stellar wind alone. The remaining two contender are: the {\em ÔX-windÕ} 
(e.g. Shang et al. 2007); and the {\em Disk-wind} (e.g. Pudritz et al. 2007). 
In the ÔX-windÕ scenario, the magnetic ÔX-pointÕ (i.e. where the stellar magnetosphere intersects the disk) 
is the point of origin of a magneto-centrifugally driven wind fueled by matter 
injected onto open field lines and flung to infinity. In this picture, magnetic forces on 
the open field lines at $\sim$0.03~AU from the central star are responsible for collimating the 
wind into a protostellar jet. Conversely, the ÔDisk-windÕ model proposes centrifugally driven 
winds launched from a magnetised disk surface, and so launching occurs not only close 
to the star but also up to a few AU out along the disk ($\sim$0.03 to 5~AU). Both models have been shown to plausibly produce observed ratios of mass ejection-accretion rates. 

\subsection{Observational challenges}\label{s:challenges}

Long standing observational difficulties in testing proposed models lie in the fact that young stars are often heavily embedded, infall and outflow kinematics are complex and confused close to the source, and the spatial and temporal scales are relatively small. For example, consider the length scales involved. The closest star forming region is in Taurus at $\sim$140\,pc, i.e. a resolution of 140~AU for typical seeing-limited observations of 1". Meanwhile, the jet acceleration and collimation zone is located at $\sim$1-40~AU above the disk-plane, thus requiring $\sim$0."1 resolution. Worse still, the central jet engine operates on scales of $<$5~AU, demanding $\sim$ 0."01 resolution. Observational difficulties persist even after the jet has travelled far from the source. Jet emission lines, which result from collisional excitation as the jet propagates into the parent cloud, mark the location of shock fronts and post-shock cooling zones which have length scales on the order of tens of AU, while jet widths are typically only $\sim$ 15 AU. Hence resolving the jet internal structure, excitation and kinematics is heavily dependent on high spatial resolution data. 

Observational approaches to overcome these challenges include focussing on more evolved, optically visible T~Tauri jets which are often traced close to the central source ($<$0."5). Meanwhile, access to the jet acceleration/collimation region, or resolving the jet width and plasma physics, requires for example the use of adaptive optics, the Hubble Space Telescope (HST) or the James Webb Space Telescope, which allow access down to scales of $<$0.$"$1. Alternatively, information can be extracted via special techniques such as spectro-astrometry, to reach scales down to $<$0.$"$01. Meanwhile, probing the central jet engine demands interferometry (VLTI/AMBER, E-Merlin, E-VLA, ALMA) which allows access to scales in the range $\sim$0.$"$1 to 0.$"$001. 

\section{Protostellar jets at high angular resolution}

\subsection{Collimation scales and initial jet velocities via adaptive optics}

Early adaptive optics imaging of the initial T~Tauri jet channel (e.g. Dougados et al. 2000) allowed the first measurements of jet collimation within 1000 AU from the star. This has proved very useful, for example, as a discriminant between the so-called `cold' and `warm' Disk-wind models. Assuming a cold disk allows enthalpy to be neglected in the jet, while assuming a warm disk (e.g. via accretion shocks or coronal heating) results in increased mass loading and divergence of field lines. The {\em observed} variation of jet radius with distance and velocity implies that the jet must be heated at its base (Ferreira et al. 2006). Meanwhile, early near infrared adaptive optics spectroscopic studies revealed a two-component velocity structure close to the base of the DG~Tau jet, with peak velocities of -220 and -100 km\,s$^{-1}$ (Pyo et al. 2003), thus prompting the question of whether there are two separate launch mechanisms at work. Subsequently, near infrared adaptive optics spectro-imaging studies were also conducted of the jets from intermediate mass counterparts to T~Tauri stars, i.e. Herbig Ae/Be stars, (e.g. Perrin et al. 2007), and suggest that the same mechanism is at work for jet generation in different mass regimes. 

\subsection{Jet plasma physics via HST}

In current star formation theory, jets/outflows from a newly forming star are believed to transport significant amounts of energy and momentum away from the region of the central source. This can have a sizeable impact on the way in which the stars form because, for example, jets may drive the injection of turbulence in the parent cloud thus delaying collapse, and thereby regulating the star formation rate. To fully understand the mechanisms underlying the physics it is necessary to know the mass outflow rate, which regulates the dynamics of the flow and is therefore the most important input parameter for any model of jet generation and propagation. 

Intrinsic emission-line luminosities can give the number of emitting particles of a given species in the observed volume, which can then lead to the gas density under an assumption of abundances.  However, this method relies on an accurate prior knowledge of reddening estimates, the excitation temperature, the ionisation state of the given species, and the filling factor, all of which bring substantial uncertainties to the calculation (e.g. Nisini et al. 2005). An alternative approach is to model emission line ratios under the assumption of a definite mechanism for the gas heating, to obtain the ionisation fraction. Although the jet electron density is readily calculated from the optical [S~II] doublet ratio, difficulties arise in pinning down a determination of the ionisation fraction which is needed to calculate the hydrogen density. However, there is as yet no general consensus on the mechanism causing the jet emission, although without a doubt the observed forbidden lines are excited collisionally. The most widely accepted explanation is that the gas is heated by internal shocks, so ionisation determinations have been made via an assumption of shock models (Hartigan et al. 1994). However, other possibilities do include ambipolar diffusion, turbulent mixing layers, and compression by jet instabilities (e.g. Lavalley-Fouquet et al. 2000 and references therein). Hence, it is preferable to obtain an estimate of the ionisation {\em without} any assumptions of heating mechanisms, for example, via ratios of optical forbidden emission line together with some simple assumptions on the gas physics which apply where low-excitation conditions exist (Bacciotti et al. 1999). 

From filtered images to slitless spectroscopy to integral-field-unit (IFU) style datacubes, HST has provided a vast wealth of information on the physics of the jet just beyond its acceleration and collimation region. Flux ratios of optical forbidden emission lines have been used to reveal jet physical parameters. Initially filtered imaging were used to derive 1D diagnostics of the first 5" of the T~Tauri jet from HH~30 (Bacciotti et al. 1999). Much later, the same jet was observed with HST/STIS at two epochs with a 2" wide slit parallel to the jet which produced an image of the jet in optical emission lines (Hartigan \& Morse 2007). Line ratio maps revealed changes in excitation conditions with time, e.g. propagation of internal jet shocks, and identification of reionisation events. Using a different approach, {\em velocity resolved} observations were obtained by stepping the HST/STIS slit across the jet to obtain an IFU-style data cube (Bacciotti et al. 2000). T~Tauri jet DG Tau was thus mapped in optical forbidden emission lines, in several velocity intervals. The excitation conditions were examined as a function of velocity, to reveal for example that electron density increases with velocity, collimation and proximity to the jet axis (Maurri et al. submitted). A similar study was conducted for the RW Aur bipolar jet (Woitas et al. 2002; 2005) and for intermediate mass Herbig Ae/Be jets e.g. LkH$\alpha$~233 (Melnikov et al. 2008). Finally, placing the HST/STIS slit {\em perpendicular} to the direction of jet propagation yields gas conditions as a function of velocity and distance from the jet axis. This revealed, for example, transverse asymmetries in electron density in the case of T~Tauri jet Th~28 (Coffey et al. 2008).

Overall, these high resolution observations within the first 100~AU (i.e. just above the accelleration \& collimation region) revealed that the jet undergoes early collimation; it presents a layered onion-like velocity and density structure with respect to the jet axis; 
velocities and densities decrease towards the jet borders; typical electron densities are upwards of 10$^3$~cm$^{-3}$; electron density is higher than the critical density of [S II] close to the jet base (i.e. $>~$2$\times$10$^3$~cm$^{-3}$); jets are partially ionised, typically anything up to 60\%; temperatures lie in the range 1 - 3 10$^4$~K; plasma conditions and velocities are often different in the jet and counter-jet, for reasons unknown; T~Tauri jet mass fluxes are on the order of $\sim$~10$^{-8}$~M$_{\odot}$~yr$^{-1}$; the ratio of jet to accretion mass flux is $\sim$~0.05 to 0.1; and, last but not least, Herbig Ae/Be jets seems to be scaled-up versions of T~Tauri jets. 

\subsection{Jet rotation via HST}

Angular momentum must be transported away from accreting systems to allow accretion to proceed, while maintaining protostar rotation at well below break-up velocity. Consequently, with the first observations of protostellar jets in the 1980s, it was proposed that jets and outflows from newly forming stars somehow extract angular momentum from their source (Section~\ref{s:models}). Until recently, backing for this theory has been hindered by observational difficulties (Section~\ref{s:challenges}). 

Excitingly, recent years have seen the first indications of jet rotation. 
The first hint came from ground-based seeing limited observations which revealed differences in radial velocity between one side of the jet axis and the other, for the HH~212 molecular jet, which were interpreted as a possible rotation signature based on agreement with the sense of disk rotation (Davis et al. 2000; Wiseman et al. 2001). However, the jet was observed far from the source allowing time for external influences to disturb the intrinsic kinematics. Independently, asymmetries in velocities indicative of rotation were identified within the first 100~AU of the DG Tau jet via the Hubble Space Telescope Imaging Spectrograph (HST/STIS) (Bacciotti et al. 2002), and were again found to be in agreement with the disk rotation sense (Testi et al. 2002). Follow-up survey observations with HST/STIS in the optical regime confirmed that systematic radial velocity asymmetries across the jet base are common in T~Tauri systems (Coffey et al. 2004). The survey was moved to the higher resolution of the near-UV, and yet again asymmetries were identified (Coffey et al. 2007). 

These findings have profound implications as the long awaiting observational support for the magneto-centrifugal class of models (Section~\ref{s:models}), and the derived toroidal velocities are in agreement with model predictions. Furthermore, these findings also act as a powerful discriminant between competing steady MHD models. Specifically, these measurements can be used to find the jet launching location on the disk plane, yielding estimates a few AU and thus supporting the Disk-wind model (Bacciotti et al. 2002; Anderson et al. 2003; Coffey et al. 2004, 2007). 

Due to these significant ramifications, our results have triggered much debate in the community. For example, it has subsequently been claimed that X-winds and Disk-winds are not mutually exclusive (Shang et al. 2007). Furthermore, debate surrounds the issue of whether these velocity asymmetries should indeed be interpreted as jet rotation. Alternative explanations include asymmetric shocking and jet precession (Cerqueira 2006). As a crucial test of jet rotation, observations must fulfill necessary criteria, namely that gradients are consistently detected in many targets, the sense of the gradient matches in both lobes of a bipolar jet, and gradients are in the same sense as the disk rotation. Controversially, in one case, namely RW Aur, the sense of jet gradient did not match that of the disk (Cabrit et al. 2006). This is a complex and highly variable system, and so other external influences may be coming into play. 

\begin{figure}
  \centerline{\includegraphics[angle=0,width=11cm]{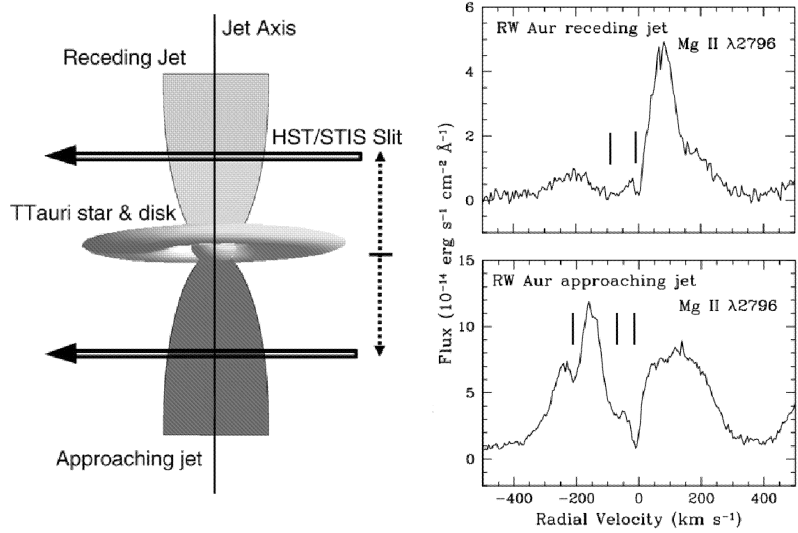}}
  \caption{Left: HST/STIS observing mode. Right: RW Aur bipolar jet observed for the first time in near-UV emission at $<$100~AU from the disk-plane (Coffey et al. in preparation). The bipolar jet emits strongly in Mg II, with a clear flux asymmetry between the two lobes. The profiles are complex, showing reflected emission from the other lobe as well as blue-shifted absorption at both low and high velocities (as marked).   \label{f:rwaur_flux}}
  \end{figure}

Following the failure of the HST/STIS power supply in August 2005, our efforts continued from the ground. Seeing limited near-infrared observations failed to resolve the atomic jet width (Coffey et al. 2010; 2011). Although, the molecular flow was resolved in younger sources and gradients detected (Coffey et al. 2011), the necessity of observing far from these embedded sources (i.e. several hundreds/thousands of AU versus 50-100 AU for HST studies) introduces uncertainty as to whether any rotation signature will maintain its integrity to such distances. Meanwhile, several ground based findings for molecular flows were reported by other groups (e.g. Codella et al. 2007; Lee et al. 2007; Chrysostomou et al. 2008; Correia et al. 2009), including CB~26 (Launhardt et al. 2009) and NGC 1333 IRAS 4A2 (Choi et al. 2010; 2011) both cases of which show agreement in jet and disk rotation sense. Meanwhile, following the repair of HST/STIS, the second phase of our HST/STIS near-UV survey got underway. Observations of 5 jet targets have been conducted, including the bipolar jet from RW Aur. Although the near-UV Mg II permitted emission jet profile is complex, Fig.~\ref{f:rwaur_flux}, this data will be a valuable contribution to help disentangle the RW Aur disk-jet rotation controversy which currently casts a shadow over the jet rotation interpretation (Coffey et al. in preparation). 

Finally, thanks to advancements in numerical simulations, the validity of the jet rotation claim can be tested in other ways. Simulations of jets launched via a time-dependent MHD Disk-wind model find that the rotation signature is indeed likely to persist to the observed distances, without fear of disruption from external influences such as asymmetric shocks (Staff et al. 2010). Complementary simulations have also been conducted of early jet propagation by our group. Initial conditions include jet rotation and plasma parameters. The output of the simulations are used to produce synthetic emission line spectra. Post-processing exactly reproduces the observing conditions, e.g. resolution and line-of-sight effects, of HST/STIS spectra which claim to present signatures of jet rotation. The synthesis spectra confirm that the rotation signature is indeed detectable, and persists to observed distances (Rubini et al. in preparation). 

\subsection{Jets from brown dwarfs via spectro-astrometry}

Excitingly, in recent years, the first observational detection was made of a jet from a brown dwarf (Whelan et al. 2005). This was made possible by exploiting the valuable technique of spectro-astronomy, which uses profile fitting to recover spatial information beyond the nominal spatial resolution of the observations. While low mass protostars lie in the range of 0.5 - 3~M$_{\odot}$, brown dwarfs occupy the mass range between the largest gas giant planets and the lowest mass protostars. Typically, they have a mass of several tens times that of Jupiter. However, by definition, brown dwarfs do not contain enough mass (i.e. $<$~0.08~M$_{\odot}$) to sustain stable hydrogen fusion, and so are in a sense 'failed stars'. Nevertheless, they now appear to follow the same formation mechanism as typical protostars via the accretion-ejection process. Moreover, the brown dwarf in question, 2MASS1207-3932, has a 5 Jupiter mass planetary companion and is surrounded by a planetary disk, just like a protostar star. The finding is highly significant as it extends the mass range for the central object in an accretion-ejection structure down to extremely low levels. This implies the ejection phenomenon is extremely robust and suggests that the same mechanism is applicable across all mass ranges from hundreds of millions of solar masses down to Jupiter sized objects. Follow-up studies continue to identify outflows for other brown dwarfs (Whelan et al. 2007; 2009). 

\section{Conclusion}

The enigma of jets in star formation is a long-standing one, observationally hindered by technological contraints, but is now reaching a crucial stage of advancement with the development of high resolution instruments. As recently as only the past decade, we have obtained an impressive wealth of information from high angular resolution facilities. These observations appear to provide the long-awaited validation of the magneto-centrifugal launching model. Furthermore, jet structure, kinematic and excitation properties have been identified and are found to be similar for different masses of central object from T~Tauri stars to Herbig Ae/Be stars, implying that the same mechanism applies for different masses. The detection of jets from brown dwarfs underlines how robust the accretion-ejection process is in operating on a vast range of mass scales. Future work requires an increased number of targets observed at high resolution, and comparison with models via numerical simulations. Observationally, the future lies with ground-based adaptive optics facilities such as GEMINI, space missions such as HST, JWST and Herschel, as well as the extremely high resolution of interferometers such as ALMA. 

\section*{Acknowledgements}

D.C. would like to thank the Indian Institute of Astrophysics in Bangalore for the invitation to speak at this conference, and for the extension of overwhelming hospitality. We acknowledge financial contribution under Italian Space Agency agreement ASI-INAF I/009/10/0. D.C. also acknowledges financial support from the Irish Research Council for Science, Engineering and Technology (IRCSET), and the Marie Curie European Reintegration Grant Scheme (ERG).

\end{document}